\documentclass{article}[12pt]

\usepackage{arxiv}

\newcommand{\that}{\widehat{\theta}}

\title{Likelihood-based solution to the Monty Hall puzzle and a related 3-prisoner paradox}
\author{Yudi Pawitan\\
Department of Medical Epidemiology and Biostatistics\\
Karolinska Institutet\\
yudi.pawitan@ki.se}

\begin{document}
\maketitle

\begin{abstract}
The Monty Hall puzzle has been solved and dissected in many ways, but always using probabilistic arguments, so it is considered a probability puzzle. In this paper the puzzle is set up as an orthodox statistical problem involving an unknown parameter, a probability model and an observation. This means we can compute a likelihood function, and the decision to switch corresponds to choosing the maximum likelihood solution. One advantage of the likelihood-based solution is that the reasoning applies to a single
game, unaffected by the future plan of the host. I also describe an earlier version of the puzzle in terms of three prisoners: two to be executed and one released. Unlike the goats and the car, these prisoners have consciousness, so they can think about exchanging punishments. When two of them do that, however, we have a paradox,
where it is advantageous for both to exchange their punishment with each other. Overall, the puzzle and the paradox are useful examples of statistical thinking, so they are excellent teaching topics.
\end{abstract}

\section{The puzzle and the paradox}
First, here is the Monty Hall puzzle:
\begin{quote}
    You are a contestant in a game show and presented with 3 closed doors. Behind one is a car, and behind the others only goats. You pick one door (let's call that Door 1), then the host \textit{will} open another door that reveals a goat. With two un-opened doors left, you are offered a switch. Should you switch from your initial choice?
\end{quote}
It is well known that you should switch: doing so will increase your
probability of winning the car from 1/3 to 2/3. The problem is
actually how to provide a convincing explanation for those who think
the chance of winning for either of the un-opened door is 50-50, so
there is no reason to switch. The puzzle, first published by Selvin
(1975ab),  was inspired by a television game show \textit{Let's Make
a Deal} originally hosted by Monty Hall. It later became famous
after appearing in Marilyn vos Savant's "Ask Marilyn" column in
Parade magazine in 1990. It generated its own literature and many
passionate arguments, \textit{even} among those who agree that you
should switch.

Martin Gardner (1959)'s article on the same puzzle in terms of three prisoners (below) was titled `Problems involving questions of probability and ambiguity.' He started by referring to the American polymath Charles Peirce's observation that `no other branch of mathematics is it so easy for experts to blunder as in probability theory.' The main problem is that some probability problems may contain subtle ambiguities. What is so clear to me -- as the statement of the Monty Hall problem above -- may contain hidden assumptions I forget to mention or even am not aware of. For example, as will be clear later, the standard solution assumes that, if you happen to pick the winning door, then the host will open a door randomly with probability 0.5. For me that is obvious so as to make the problem solvable and the solution neat. If we explicitly drop this assumption, then there is no neat solution; but our discussion is not going to that direction.

Instead, we head to a more interesting paradox. In Martin Gardner's version there are three prisoners A, B and C, two of whom will be executed and one released. Being released is like winning a car. Suppose A asks the guard: since for sure either B or C will be executed, there is no harm in telling me which one. Suppose the guard says C will be executed. Should A asks to switch his punishment with B's? With the same logic as in the Monty Hall problem, it must be yes, A should switch with B. But, what if B asks the same question to the guard, which of A or C will be executed, and the guard also answers C? Then it seems also advantageous for B to switch with A. Now we have a paradox: how can it be advantageous for both A and B to switch?

We can make another version of the paradox: Suppose A just \textit{heard the breaking news} that C has been executed (no guard is involved). Should A ask to switch his punishment with B? What logic should apply here? If the same as before, then he should switch. But then, the same logic applies to B, and we arrive at the same paradox. If not the same, then having the guard to answer the questions matters. But why does it matter how A finds out that C is (to be) executed?

In probability-based reasoning we are considering:  what is the
probability of winning if you stay with your initial choice vs if
you switch, This is relatively easy -- hard enough for some people
-- with the Monty Hall puzzle, but it gets really challenging if we
want to explain the 3-prisoners paradox. So, instead, the puzzle and
the paradox will be set up as an orthodox statistical problem
involving an unknown parameter, a probability model and an
observation. This means we can compute a likelihood function, and
the decision to switch corresponds to choosing the maximum
likelihood solution. In the Monty Hall puzzle, one advantage of the
likelihood-based solution is that the reasoning applies to a single
game, unaffected by the future plan of the host. In addition, the
likelihood construction will show explicitly all the technical
assumptions made in the game.

\section{Likelihood-based solutions}
\subsubsection*{Monty Hall problem}
Let $\theta$ be the location of the car, which is completely unknown to you. Your choice is called Door 1; for you the car could be anywhere, but of course the host knows where it is. Let $y$ be the door opened by the host. Since he has to avoid opening the prized door, $y$ is affected by both $\theta$ and your choice. The probabilities of $y$ under each $\theta$ are given in this table:
\begin{center}
    \begin{tabular}{rcccc}
     & $\theta=1$ & $\theta=2$ & $\theta=3$ & $\that$ \\
    \hline
    $y=1$ & 0 & 0 & 0 & -\\
    2 & 0.5&0 & {1}& 3\\
    3 & 0.5& {1} & 0& 2\\
    \hline
    Total & 1 & 1 & 1
\end{tabular}
\end{center}
For example, it is clear that $y$ cannot be 1, because the host cannot open your door (duh!), so it has zero probability under any $\theta$. If your Door 1 is the winning door ($\theta=1$), the host is \textit{assumed} to choose randomly between Doors 2 and 3. If $\theta=2$, he can only choose Door 3. Finally, if $\theta=3$, he can only choose Door 2. So, the `data' in this game is the choice of the host. Reading the table row-wise gives the likelihood function of $\theta$ for each $y$, so the maximum likelihood estimate is obvious:
\begin{center}
If $y=2$, then $\that=3$ (so you should switch from 1 to 3.) \\
If $y=3$, then $\that=2$ (switch from 1 to 2.)
\end{center}
The increase in the likelihood of winning by switching is actually only 2 folds, so it is not enormous. But the increase in prize from a goat to a car is enormous, so a switch is warranted.

In this likelihood formulation, the problem is a classic statistical problem with  data following a model indexed by an unknown parameter. For orthodox non-Bayesians, there is no need to assume that the car is randomly located, so no probability is involved on $\theta$; it is enough to assume that you are completely ignorant about it. From the table, it is clear also that, when $\theta=1$, the host does not need to randomize with probability 0.5; he can do that any probability at all and you will not reduce your likelihood of winning by switching.

\subsection*{Probability or likelihood?}\label{sec:3}

Previous solutions of the Monty Hall puzzle are typically given in terms of probability. Why bother with the likelihood? Imagine a forgetful host in the Monty Hall game: he forgets which door has the car, so he opens a door at random. If it reveals a goat, then the game can go on; if it reveals the car, then he makes excuses, and the game is cancelled and nobody wins. Suppose in your particular game, a goat is revealed. Is it still better to switch? As before, let's call your chosen door as Door 1. Define the data $y$ as the \textit{un-opened door} (other than yours) if the opened one is a goat (so the game is on); if the opened one is a car, then  set $y\equiv 4$ (and the game is off). The probability table is now:
\begin{center}
    \begin{tabular}{rcccc}
     & $\theta=1$ & $\theta=2$ & $\theta=3$ & $\that$ \\
    \hline
     $y=2$ & 0.5 & 0.5 & 0 & \{1,2\}\\
    $3$ & 0.5 & 0 & 0.5 & \{1,3\}\\
    4 & 0& 0.5 & 0.5& -\\
    \hline
        Total & 1 & 1 & 1
\end{tabular}
\end{center}
So, when $y=2$ or 3, and the game still on, your likelihood of winning the car with your original or the other unopened door is equal, and there is no benefit of switching. Comparing with the original version, this means that the evidence of an open door revealing a goat is not sufficient to say that switching is beneficial. We must also know whether it was intentional or accidental. But, how about the future plan? Should it also affect your reasoning? In technical probability reasoning it should also matter, because probability is not meant to apply for a specific game.

Say in the current game the host \textit{intentionally} opens a door with a goat, but in the future he \textit{plans} to open a door randomly. What logic applies to the current game? Non-Bayesians certainly cannot apply the orthodox probability-based argument to the current game. But, \emph{the likelihood-based reasoning still applies to the current game, unaffected by the future plan of the host.}

\subsection*{3-prisoner paradox: guard involved}
The likelihood-based reasoning also helps solve the 3-prisoner paradox.
Let  $\theta$ be the identity of the prisoner to be released; to avoid
confusion, set $\theta=$ 1, 2 or 3 for the three prisoners A, B and C. Let $A_1$ be the guard's answer to prisoner A, also denote the answer by 1, 2 and 3. Then we have indeed the same probability table as for the game show:
\begin{center}
    \begin{tabular}{rcccc}
     & $\theta=1$ & $\theta=2$ & $\theta=3$ & $\that$ \\
    \hline
    $A_1=1$ & 0 & 0 & 0 & -\\
    2 & 0.5&0 & {1}&3\\
    3 & 0.5& {1} & 0&2\\
    \hline
    Total & 1 & 1 & 1
\end{tabular}
\end{center}
Here, the guard cannot answer $1$ to prisoner A, and he will randomize whenever possible. So, we get the same maximum-likelihood estimate as above, leading to switching as a good strategy.

When B asks the same question, let $A_2$ be the guard's answer. The joint distribution of $(A_1,A_2)$ is as follows. It can be derived under the same requirements: the guard cannot tell the questioner that he would be executed, and he must randomize whenever possible.
\begin{center}
    \begin{tabular}{rccc}
    & $A_1=1$ & $A_1=2$ & $A_1=3$ \\
    \hline
   $\theta=1, A_2=1$ & 0 & 0 & 0 \\
    $A_2=2$ & 0&0 &0\\
    $A_2=3$ & 0& 0.5 & 0.5\\
    \hline
       $\theta=2, A_2=1$ & 0 & 0 & 0.5 \\
    $A_2=2$ & 0&0 &0\\
    $A_2=3$ & 0& 0 & 0.5\\
    \hline
       $\theta=3, A_2=1$ & 0 & 1 & 0 \\
    $A_2=2$ & 0&0 &0\\
    $A_2=3$ & 0& 0 & 0\\
    \hline
\end{tabular}
\end{center}
Keeping only the non-trivial scenarios, the table can be simplified to
\begin{center}
    \begin{tabular}{rcccccc}
 & $\theta=1$ & $\theta=2$ & $\theta=3$ & $\that$ & Better for A & Better for B \\
    \hline
  $(A_1,A_2)$=(2,1) & 0 & 0 & 1 & 3 & Switch with C& Switch with C\\
    (2,3) & 0.5&0 & 0 & 1 & Don't switch! & Switch with A\\
    (3,1) & 0& 0.5 & 0 & 2 & Switch with B & Don't switch!\\
    (3,3) &0.5& 0.5& 0 &  \{1,2\} & None& None\\
    \hline
        Total & 1 & 1 & 1
\end{tabular}
\end{center}
Based on the relevant outcome of the story $(A_1=3,A_2=3)$, the prisoners A and B actually have equal likelihood of being released. So, a switch confers no advantage to either side, and there is no paradox.

How can the paradox appear? Let's consider who has access to what information. If A only knows $A_1=3$, but does not know nor presume the existence of $A_2$, then his reasoning is incomplete. Similarly, B only knows $A_2$. So, the paradox appears because of incomplete information by each side. An external agent -- e.g. the guard -- who knows both answers $A_1$ and $A_2$ can see there is no advantage in switching.

Suppose, both A and B know that each has asked the relevant question, but A only knows his own answer $A_1$ and B only knows $A_2$. Furthermore, assume that a switch can only happen by mutual consent. Suppose $A_1=3$, then, for A, a switch is advantageous only when $A_2=1$, but neutral when $A_2=3$. But $A_2=1$ means B is told that A will be executed, so he will not be willing to switch with A. The same logic applies of B. So, when both sides are willing, they know the switch must be neutral, and there is no paradox.

\subsection*{3-prisoner paradox: news version}

How about the news break version? As before,let $\theta$ be the identity of the prisoner to be released. Now, consider the data $y$ is the first prisoner reported to be executed, assuming that execution is in random order. Then we can derive the probabilities under different $\theta$'s.
\begin{center}
    \begin{tabular}{rcccc}
     & $\theta=1$ & $\theta=2$ & $\theta=3$ & $\that$ \\
    \hline
    $y=1$ & 0 & 0.5 & 0.5 & \{2,3\}\\
    2 & 0.5&0 & 0.5& \{1,3\}\\
    3 & 0.5&0.5 & 0&\{1,2\} \\
    \hline
        Total & 1 & 1 & 1
\end{tabular}
\end{center}
So, when $y=3$, A and B have equal likelihood of being released. In other words, when both A and B heard that C has been executed, there is no advantage for either of them to switch.

Why does the explanation look so trivial? Well, we can make it more complicated. The fact that A heard the news means that he was not the first to be executed. Suppose A was in fact told that, if he is to be executed (which he might not), then he will not be the first. Does that affect he should react to the news? The probability table becomes:
\begin{center}
    \begin{tabular}{rcccc}
     & $\theta=1$ & $\theta=2$ & $\theta=3$ & $\that$ \\
    \hline
     $y=1$ & 0 & 0 & 0 & -\\
    $2$ & 0.5 & 0 & 1 & 3\\
    3 & 0.5& 1 & 0& 2\\
    \hline
        Total & 1 & 1 & 1
\end{tabular}
\end{center}
Hey, this looks familiar! Of course, it is exactly the same table as for the Monty Hall puzzle above. So, yes, in this case the news is informative: A should indeed ask to switch with B. But what did they tell B? Suppose B knows that A is told that A will not be the first to be executed, but B himself is not told that. How should he react to the news? From the table, when $y=3$, then $\that=2$, so B should not switch with A. If both are told -- and both know this -- that they won't be the first to be executed, then the table becomes:
\begin{center}
    \begin{tabular}{rcccc}
     & $\theta=1$ & $\theta=2$ & $\theta=3$ & $\that$ \\
    \hline
     $y=1$ & 0 & 0 & 0 & -\\
    $2$ & 0 & 0 & 0 & -\\
    3 & 1& 1 & 0& \{1,2\}\\
    \hline
\end{tabular}
\end{center}
So, when $y=3$, we are back to the situation of neutral switch.

\section{Conclusion}

Not surprisingly, different decisions by the player in the Monty
Hall  problem or by the prisoners depend on different model
setup/assumptions and available data. The same data, e.g. an open
Door 3 revealing a goat, can be interpreted differently depending on
the setup. The likelihood-based calculation requires explicit
assumptions, so it clarifies them. We also compare the probability-
vs likelihood-based reasoning; the advantage of the latter is its
applicability to a single game, unaffected by future intention or plans. Gill (2011) provided a clear discussion of the assumptions associated with the standard probability-based solution of the puzzle as well as a two-person game that shows switching as a minimax solution. The model formulation here agrees with Gill that the Monty Hall problem is not a probability puzzle; he considered it a mathematical modelling problem. Here I have phrased it as an orthodox statistical problem so it -- and the related paradox -- can be a useful example in likelihood-based modelling.

\section{References}
\begin{description}
\item Gardner, Martin (1959). "Mathematical Games: Problems involving questions of probability and ambiguity". \textit{Scientific American}, 201 (4): 174–182.
\item Gill, Richard (2011). The Monty Hall problem is not a probability puzzle* (It's a challenge in mathematical modelling). \textit{Statistica Neerlandica}, 65 (1), 58–71.
\item Selvin, Steve (1975a). "A problem in probability". \textit{American Statistician}, 29 (1), 67–71.
\item Selvin, Steve (1975b). "On the Monty Hall problem". \textit{American Statistician}, 29 (3), 134.

\end{description}
\end{document}